\documentclass[useAMS,usenatbib]{mn2e}

\usepackage{times} 
\usepackage{graphicx}
\usepackage{rotating}
\usepackage{epsfig}

\newcounter{subfigure}

\newcounter{species} 
\def\ion#1#2{\setcounter{species}{#2}#1$\;${\scriptsize\Roman{species}}\relax}

\newcommand{\Lya}{\hbox{{\rm Ly}\kern 0.1em$\alpha$}}
\newcommand{\Lyb}{\hbox{{\rm Ly}\kern 0.1em$\beta$}}
\newcommand{\Lyg}{\hbox{{\rm Ly}\kern 0.1em$\gamma$}}
\newcommand{\Lyd}{\hbox{{\rm Ly}\kern 0.1em$\delta$}}
\newcommand{\Lye}{\hbox{{\rm Ly}\kern 0.1em$\epsilon$}}
\newcommand{\CIVdblt}{{\rm C}\kern0.1em{\sc iv}~$\lambda\lambda 1548, 1551$}
\newcommand{\NVdblt}{\hbox{{\rm N}\kern 0.1em{\sc v}~$\lambda\lambda 1239, 1243$}}
\newcommand{\OVIdblt}{{\rm O}\kern 0.1em{\sc vi}~$\lambda \lambda 1032, 1038$}
\newcommand{\MgIIdblt}{{\rm Mg}\kern 0.1em{\sc ii}~$\lambda\lambda 2796, 2803$}
\newcommand{\AlIIIdblt}{{\rm Al}\kern0.1em{\sc iii}~$\lambda\lambda 1855, 1863$}
\newcommand{\SiIVdblt}{{\rm Si}\kern 0.1em{\sc iv}~$\lambda\lambda 1393, 1402$}

\newcommand{\kms}{\hbox{km~s$^{-1}$}}

\newcommand{\etal}{et~al.}

\title[Catalog of Quasar Absorption Lines]{A Catalog of Absorption
       Lines in Eight HST/STIS E230M $1.0 < z < 1.7$ Quasar Spectra
       \thanks{Based on observations obtained with the NASA/ESA Hubble
       Space Telescope, which is operated by the Space Telescope
       Science Institute (STScI) for the Association of Universities
       for Research in Astronomy, Inc., under NASA contract
       NAS5D26555.} }
\author[N. Milutinovi\'c {\etal}]{
        N.~Milutinovi\'c,$^{1,2}$
        T.~Misawa,$^{1}$
        R.~S. Lynch,$^{1,3}$
        J.~R. Masiero,$^{1,4}$
        C.~Palma,$^{1}$
        J.~C. Charlton,$^{1}$
        \newauthor
        D.~Kirkman,$^{5}$
        S.~Bockenhauer$^{5}$ and
        D.~Tytler$^{5}$ \\
        $^{1}$ Department of Astronomy \& Astrophysics, 525 Davey Lab,
         Pennsylvania State University, University Park, PA 16802 \\
        $^{2}$ Department of Physics \& Astronomy, University of
         Victoria, Elliott Building, 3800 Finnerty Rd, Victoria, BC,
         V8P 5C2 Canada \\
        $^{3}$ Department of Astronomy, P.O Box 400325, University of
         Virginia, Charlottesville, VA 22904 \\
        $^{4}$ Institute for Astronomy, University of Hawaii 2680
         Woodlawn Dr, Honolulu, HI 96822 \\
        $^{5}$ Center for Astrophysics and Space Sciences, University
        of California San Diego, MS 0424, La Jolla, CA 92093-0424}

\begin{document}

\date{*A complete version can be obtained at
http://www.astro.psu.edu/users/misawa/pub/Paper/qalcat.pdf.gz}

\pagerange{\pageref{firstpage}--\pageref{lastpage}} \pubyear{2007}

\maketitle

\label{firstpage}

\begin{abstract}
We have produced a catalog of line identifications and equivalent
width measurements for all absorption features in eight ultraviolet
echelle quasar spectra. These spectra were selected as having the
highest signal-to-noise among the {\it HST}/STIS spectra obtained with
the E230M grating.  We identify 56 metal-line systems toward the eight
quasars, and present plots of detected transitions, aligned in
velocity-space.  We found that about 1/4 - 1/3 of the features in the
{\Lya} forest region, redward of the incidence of the {\Lyb} forest,
are metal lines. High ionization transitions are common. We see both
\ion{O}{6} and \ion{C}{4} in 88 -- 90\%\ of the metal-line systems for
which the spectra cover the expected wavelength. \ion{Si}{3} is seen
in 58\%, while low ionization absorption in \ion{C}{2}, \ion{Si}{2},
and/or \ion{Al}{2} is detected in 50\%\ of the systems for which they
are covered. This catalog will facilitate future studies of the {\Lya}
forest and of metal-line systems of various types.
\end{abstract}


\section{Introduction}
The most important chemical transitions for quasar absorption line
systems at low redshift (e.g., $z$$<$1) still lay in the ultra-violet
spectral range. A limited number of the most UV-bright quasars have
been observed with the Hubble Space Telescope ({\it HST})/ Space
Telescope Imaging Spectrograph (STIS), which has provided a detailed
view of absorption systems along the lines of sight.  These data are
available from the Multimission Archive at Space Telescope (MAST).
However, in order to compile statistics of absorbers or to study
particular systems, it is necessary to identify the spectral features.
This can be complicated, particularly in the {\Lya} forest region, and
when the available spectral coverage is limited.

In the course of our studies of the {\Lya} forest and metal-line
systems we have studied in detail eight of the highest quality {\it
HST}/STIS spectra, obtained with the E230M grating.  In this paper we
present these spectra with a list of $5~\sigma$ absorption features
and our suggested line identifications.  In particular, metal-line
system identifications are necessary in order to remove contaminants
from the {\Lya} forest.  In a recent paper \citep{kir07}, we measured
statistics of absorption in the {\Lya} forest at $0 < z < 1.6$ using
74 low resolution {\it HST}/FOS quasar spectra.  As a check, and
particularly to assess our ability to separate metal lines from the
forest, we also used the available higher resolution E230M {\it
HST}/STIS spectra.  In the present paper, we present the relevant data
and line identifications which were used in that paper.  We expect
that this catalog will facilitate future studies of metal-line
absorption systems as well as of the {\Lya} forest.

Our catalog includes damped {\Lya} absorbers (DLAs), Lyman limit
systems, and {\Lya} forest clouds.  The DLA and Lyman limit systems
are expected to have strong \ion{Mg}{2} absorption \citep{rao00,
archive1}, while the {\Lya} forest clouds may have weak \ion{Mg}{2}
and other low ionization transitions detected (such as \ion{Si}{2} and
\ion{C}{2}) \citep{nar05}.  Some {\Lya} forest clouds have detected
\ion{C}{4} and/or \ion{O}{6}, without associated low ionization
absorption \citep{tripp,savage,mil06}.  The catalog that we will
present will allow for future studies of the various types of
absorbers.

In \S~\ref{sec:data} we describe {\it HST}/STIS quasar spectra and our
reduction and continuum fitting procedures.  We also outline our line
detection and measurement algorithms and line identification
strategies.  The normalized echelle spectra are presented in
\S~\ref{sec:results}, along with a brief description of metal-line
systems detected toward each quasar.  System plots, showing all
detected transitions for each system, are also presented there.  In
\S~\ref{sec:summary} we present a general summary of the systems
presented in this catalog.

\section{Data and Methods}\label{sec:data}

We selected the eight $z>1$ quasars observed with the {\it HST}/STIS
E230M grating which had a signal to noise ratio, $S/N > 5$, over a
reasonable fraction of the spectrum.  These {\it HST}/STIS E230M
spectra have a resolution $R=30,000$, with two pixels per resolution
element. Our sample is biased towards quasars that have Lyman limit
systems, since most of the STIS observations were conducted in order
to study those particular known systems in detail.

Several settings are possible covering different wavelength ranges.
For seven of the eight quasars in our sample, $2280$ - $3110$~{\AA} is
covered.  The quasars and the relevant observational details are
listed in Table \ref{tab:tab1}.  Specifically, this table lists quasar
redshifts, wavelength coverage, $S/N$ per pixel at 2350~{\AA} and
2750~{\AA}, primary investigator for the original observation, and
proposal ID.

The data were reduced using the STIS pipeline \citep{bro02}.  They
were combined by simple weighting by exposure time (as in
\citet{nar05}), rather than by, e.g., inverse variance methods that
are often employed.  A bias could be introduced by the latter method,
since pixels with smaller counts would be weighted more heavily.  This
bias is significant only in cases such as these STIS spectra, where
individual exposures have very small $S/N$, and is not typically
important for reduction of ground-based high resolution echelle
spectra.

Wavelengths were corrected to the heliocentric reference frame.  When
the same quasar was observed multiple times, there was often a small
shift in wavelength between the spectra, due to the intentional
shifting of the echelle angle of the instrument.  In this case, we
chose not to smooth the data by interpolating, but instead chose
wavelength bins from one exposure and combined the flux from other
exposures into the nearest bin.  This results in a slightly decreased
effective resolution. Continuum fits used the standard IRAF SFIT
task.\footnote{IRAF is distributed by the National Optical Astronomy
Observatories, which are operated by AURA, Inc., under contract to the
National Science Foundation.}

Features were objectively identified by searching for an unresolved
line at each pixel (as employed in \citet{sch93}), and applying a
$5\sigma$ criterion for detection.  However, we found by inspection
that some of these formal detections are not likely to be real.  In
some cases, only 1 pixel had significant absorption, and in other
cases it appeared that correlated noise biased the measurement.
Features that are broad and shallow can be $>$ 5$\sigma$ and yet look
unconvincing because they are sensitive to the continuum level. Such
spurious features, estimated to be 10\% of the total number of
detections, were eliminated from consideration. We define a feature as
all the pixels around a 5$\sigma$ detection that have not recover to
the continuum after smoothing the spectra using an equation on p.56 of
\citet{sch93}. This definition means that many features are clearly
blends of well separated lines, and hence a feature can have more than
one identification, each of which may be secure because they refer to
different blended lines. Wavelength that we give for a feature is the
flux-weighted central value, the mean of the wavelengths where the
smoothed flux drops below the continuum, with weighting for the
fraction of the photons that are absorbed at each pixel.

We attempted to identify all features in each of the quasar spectrum,
and the transitions that were identified one or more times are listed
with their vacuum wavelengths and oscillator strengths in
Table~\ref{tab:tab2}.  For our line identification we used the
following procedure:

\begin{enumerate}
  \item{Examine the spectrum at the expected positions of possible
        Galactic lines, and mark all possible detections.}
  \item{Search for the strong resonant doublet transitions:
        {\MgIIdblt}, {\SiIVdblt}, {\CIVdblt}, {\NVdblt}, and
        {\OVIdblt}.  In addition to these, we also search for low
        ionization gas through the combination \ion{C}{2}~$\lambda
        1335$ and \ion{Si}{2}~$\lambda 1260$, which can be used in
        place of {\MgIIdblt} when the latter is not covered
        \citep{nar05,mil06}.}
  \item{Check positions of other transitions that could be detected
        for systems found through the doublet search, including the
        Lyman series lines.  If a system is ambiguous with the doublet
        alone, {\Lya} and other Lyman series lines were used to assess
        its reality.}
  \item{Having completed these steps, all features that remain without
        plausible identifications are considered to be Lyman series
        lines.}
  \item{Begin by assuming that the highest order Lyman series line is
        the correct identification and search for all of the stronger
        (lower order) Lyman series lines corresponding to the same
        redshift.  If reasonable ratios are found, mark these as
        possible identifications.  Otherwise continue assessing the
        possible identifications of the feature in question until
        reaching {\Lya}.}
  \item{If no identification, including a {\Lya} identification, is
        plausible the feature is left unidentified.}
  \item{If more than one identification is reasonable (or if there is
        likely to be a blend for which more than one system
        contributes significantly), this is noted.}
\end{enumerate}

\section{Results}\label{sec:results}

Figure~1a is a sample of a portion of the normalized spectrum of
PG$0117+213$, shown in the printed version.  We present all other
normalized quasar spectra in Figures~1b--8d (see the complete
version), which are available electronically at
http://www.blackwellpublishing.com/products/.  Similarly, our listings
of line identifications are presented electronically in
Table~\ref{tab:tab3}, with some sample listings appearing in the
printed version.  In this table, quasars are ordered by right
ascension and $5\sigma$ features by increasing wavelength within a
given quasar entry. In Table~\ref{tab:tab3} we list, for each
$5\sigma$ feature that we did not reject as unconvincing, the observed
vacuum wavelength ($\lambda_{obs}$), the observed frame equivalent
width and error ($W_{obs}$ and $\sigma$($W_{obs}$)), and the
significance level ($S$; observed equivalent width/ error in
equivalent width).  It also gives our favored line identification for
each feature (transition and redshift) along with notes that indicate
alternative and/or additional identifications.  If the identification
cannot be confirmed with another line, or is uncertain for some other
reason, a "?" is listed after the preferred transition ID.  We do not
give column densities since there are best done on a system by system
basis using velocity plots to ensure that the columns for different
ions are representative of the same gas.

The following subsections present the eight quasars along with
discussion of noteworthy metal-line systems found in their spectra.
Figure~9 presents an example ``system plot'' for the $z=0.5764$ system
toward PG$0117+213$.  The transitions detected at $5\sigma$ for all of
the metal-line systems are plotted in velocity space in the electronic
versions of Figures~9--64 (see the complete version).  The quasar
emission redshifts, taken from the NASA/IPAC Extragalactic Database,
are listed in the heading of each subsection.

\subsection{PG0117+213 ($z_{em} = 1.493$)\label{sec:pg0117}}

This quasar was studied as part of the Quasar Absorption Line Key
Project, using G270H grating of {\it HST}/Faint Object Spectrograph
(FOS), with resolution of $R$=1300. \citet{jan98} found possible
metal-line systems at $z=0.5766$, 0.9400, 0.9676, 1.0724, 1.3389,
1.3426 and 1.3868.  They also found tentative suggestions of
metal-line systems at $z=1.3256$ and 1.4952.  \ion{Mg}{2} is detected
in a Keck/HIRES spectrum at $z=0.5764$, 0.7290, 1.0480, 1.3250 and
1.3430 \citep{chu01,chu99}, and these five systems were modeled
(including constraints from this STIS coverage) by \citet{mas05}.

\citet{jan98} observed this quasar with {\it HST}/STIS to facilitate a
statistical study of the {\Lya} forest.  This spectrum covers the
{\Lya} forest over almost all of its wavelength coverage, blueward of
$3031$~{\AA}.  The {\Lyb} line is covered blueward of $2558$~{\AA},
{\Lyg} blueward of $2423$~{\AA}, and {\Lyd} blueward of $2368$~{\AA}.
There is also limited redshift coverage of higher order Lyman series
lines.  Along this line of sight are five \ion{Mg}{2} absorbers.  One
of them is a DLA, two are strong \ion{Mg}{2} absorbers, and the other
two are multiple-cloud, weak \ion{Mg}{2} absorbers.  Three certain,
and two possible \ion{O}{6} systems are also detected.  The metal-line
systems toward this quasar are plotted in Figures~9--18.

\begin{description}

\item[\it z = 0.5764. ---] The $z=0.5764$ system is a strong
\ion{Mg}{2} absorber.  Based upon only a low resolution {\it HST}/FOS
spectrum, obtained in spectropolarimetry mode \citep{kor98}, it is not
clear whether this is a DLA or whether it has multiple undamped
components \citep{rao00}.  In the {\it HST}/STIS spectrum, {\CIVdblt}
are detected, but both transitions suffer from blending.  The
\ion{C}{4}~$\lambda 1548$ is blended in its blue component with {\Lyg}
from a system at $z=1.5088$.  An unidentified blend appears on the red
wing of \ion{C}{4}~$\lambda 1550$ and its blueward component also has
a small contribution from \ion{Si}{2}~$\lambda 1193$ at $z=1.0480$.
{\AlIIIdblt} is detected, but the blueward member has a blend to its
blue, probably with {\Lya}.  Strong \ion{Al}{2}~$\lambda 1671$ is
detected, as is \ion{Fe}{2}~$\lambda 1608$.  The strong feature
redward of \ion{Al}{2}~$\lambda 1671$ is probably {\Lya} at
$z=1.1671$. In addition to \ion{Mg}{2} and \ion{Mg}{1}, \ion{Ca}{2}
and \ion{Ti}{2} were detected in the Keck/HIRES spectrum
\citep{mas05}.

\item[\it z = 0.7290. ---] The unusual multiple-cloud, weak
\ion{Mg}{2} absorber at $z=0.7290$ is detected only in
\ion{Al}{2}~$\lambda 1671$ and \ion{C}{2}~$\lambda 1335$ in the {\it
HST}/STIS spectrum.  {\CIVdblt} is covered at a high sensitivity, yet
it is not detected to a $3\sigma$ rest-frame equivalent width limit of
$0.01$~{\AA}. This makes it ``\ion{C}{4}--deficient'', possibly
indicative of a low level of star formation so that a corona would be
weak or absent in its very red, barred spiral galaxy host
\citep{mas05}.

\item[\it z = 1.0480. ---] The $z=1.0480$ system was found not to have
metal lines detected in the low resolution {\it HST}/FOS spectrum.
The detections of \ion{Si}{2}, \ion{C}{2}, \ion{Si}{3}, {\SiIVdblt} in
the {\it HST}/STIS spectrum are consistent with those limits.
Unfortunately, the STIS spectrum does not cover {\CIVdblt}, which
would provide useful constraints on the system's physical
conditions. Based upon photoionization modeling \citep{mas05} it would
appear that there is a significant blend with \ion{Si}{3}~$\lambda
1207$, which is most likely {\Lya}.

\item[\it z = 1.3250. ---] The $z=1.3250$ system is just above the
borderline to qualify as a strong \ion{Mg}{2} absorber.  It has six
weak components in {\MgIIdblt} spread over $\sim 250$~{\kms}
\citep{chu01}.  There is a partial Lyman limit system detected in an
{\it HST}/FOS spectrum \citep{jan98}.  In the {\it HST}/STIS spectrum
\ion{Si}{2} and \ion{C}{2} are detected from the strongest four of
these six components.  \ion{Si}{3}~$\lambda$1207 is also detected over
the full velocity range, but it suffers from a blend with Galactic
\ion{Mg}{2}~$\lambda 2803$ to the blue.  There may also be blends
redward of this, based upon the photoionization models \citep{mas05},
but we were unable to provide identifications besides possible {\Lya}.
Unfortunately, {\CIVdblt}, which was detected in the {\it HST}/FOS
spectrum \citep{jan98}, is not covered in the STIS spectrum. There is
a possible weak detection of \ion{N}{5}~$\lambda$1239 but it cannot be
confirmed using \ion{N}{5}~$\lambda$1243 because it is in a noisy part
of the spectrum. A strong {\OVIdblt} is detected, but
\ion{O}{6}~$\lambda 1032$ is blended to the red with {\Lyb} from the
system at $z=1.3390$, and to the blue with possible {\Lya}.  Also,
\ion{O}{6}~$\lambda 1038$ must have a blend, at least on the red wing
of its strong component, since that profile is too strong relative to
\ion{O}{6}~$\lambda 1032$ at that velocity.  This blend could be
\ion{O}{6}~$\lambda 1032$ from the $z=1.3390$ system.  If so, these
systems would be ``line-locked''.

\item[\it z = 1.3390. ---] We find the evidence for metal lines in the
$z=1.3390$ system to be inconclusive.  {\Lya} is not detected in the
redward portion range of the possible \ion{C}{3}~$\lambda 977$
feature, so not all of that absorption can be \ion{C}{3} at this
redshift.  Two components of \ion{Si}{3}~$\lambda 1207$ are possible,
but they do not align perfectly with the minima of the \ion{C}{3}
profile and cannot be confirmed in any other way.  Also, {\OVIdblt}
may be detected.  The alignment is reasonably good, though the feature
redward of the \ion{O}{6}~$\lambda 1032$ component is likely to be
\ion{O}{6}~$\lambda 1038$ from the system at $z=1.3250$.

\item[\it z = 1.3430. ---] The $z=1.3430$ system is also just at the
border between strong and multiple-cloud weak \ion{Mg}{2} absorption.
Like the $z=1.3250$ system, it has a partial Lyman limit break in the
{\it HST}/FOS spectrum \citep{jan98}.  {\Lya} and {\Lyb} are detected
in the {\it HST}/STIS spectrum along with \ion{C}{2}, \ion{Si}{2},
\ion{N}{2}, \ion{C}{3}.  There is a feature at the expected position
of \ion{N}{3}~$\lambda 998$, which appears to be too strong relative
to the other transitions in this system \citep{mas05}.  We believe
this feature could be {\Lya} at $z=0.9076$.  \ion{C}{4} is not covered
in the STIS spectrum, but strong \ion{C}{4} absorption is detected in
the FOS spectrum.  {\NVdblt} appears to be detected, but the
\ion{N}{5}~$\lambda 1239$ transition is blended, probably with {\Lya}
at $z=1.3866$.

\item[\it z = 1.4242. ---] There is an \ion{O}{6} system, with
detected {\Lya}, {\Lyb}, and {\Lyg} at $z=1.4242$.

\item[\it z = 1.4463. ---] Another weak, broad \ion{O}{6} doublet at
$z=1.4463$ is also detected in \ion{C}{3} and \ion{Si}{3}, as well as
in {\Lya}.  The feature in the \ion{Si}{3}~$\lambda 1207$ panel, at
$\sim 40$~{\kms}, is at least partially \ion{Si}{2}~$\lambda 1260$ at
$z=1.3430$ (as is the stronger feature to its red).

\item[\it z = 1.4478. ---] At $z=1.4478$ the Lyman series is detected
down to \Lye, which is the last member covered in the {\it HST}/STIS
spectrum.  There is a possible \ion{O}{6} doublet at this redshift,
but the feature at the position of \ion{O}{6}~$\lambda 1032$
transition does not match the profile of that at the position of
\ion{O}{6}~$\lambda 1038$.

\item[\it z = 1.5088. ---] The \ion{O}{6} system at $z=1.5088$ is
another example of associated \ion{O}{6} absorption that appears
slightly redward of the quasar emission redshift.  It is detected in
{\Lya} and {\Lyb}.  The {\Lyb} shows the same components as
\ion{O}{6}, but clearly has a blend to the red, probably with {\Lya}.
{\Lyg} is heavily blended with \ion{C}{4}~$\lambda 1548$ from the
$z=0.5764$ system.  \ion{C}{3}~$\lambda 977$ is also detected in three
components, though the blueward one is affected by a data defect.

\end{description}

\subsection{HE0515-4414 ($z_{em}=1.713$)\label{sec:he0515}}

This quasar was observed with {\it HST}/STIS by \citet{rei03} in order
to study a sub-DLA at $z=1.15$, for which they also have optical
coverage.  Their study of this system, particularly focused on
molecular hydrogen, is published as \citet{rei03}.  In addition,
\citet{rei03} and \citet{lev03} report on six \ion{O}{6} systems along
the line of sight (at $z=1.385$, 1.416, 1.602, 1.674 and 1.697), based
upon the {\it HST}/STIS spectrum in conjunction with VLT/UVES data.
{\Lya} is covered over the entire {\it HST}/STIS spectrum, {\Lyb} up
to $2784$~{\AA}, and {\Lyg} is covered up to $2651$~{\AA}.  There is
more limited coverage of all other Lyman series transitions down to
the Lyman limit of the quasar.

The {\it HST}/STIS spectrum is dominated by the sub-DLA at $z=1.15$
and its many detected metal and molecular hydrogen transitions.  In
addition, there is one \ion{C}{4} system, four definite \ion{O}{6}
systems, and three possible \ion{O}{6} systems.  A couple of these
\ion{O}{6} systems are associated.  System plots for this quasar are
given in Figures~19--27.

\begin{description}

\item[\it z = 0.9406. ---] A two component \ion{C}{4} doublet is
detected at $z=0.9406$, which also has {\Lya} and \ion{Si}{3}
detected.  \ion{Si}{4} is also probably detected in the redward
component, but this can't be confirmed because \ion{Si}{4}~$\lambda
1394$ is blended with an unknown line, probably {\Lya}.

\item[\it z = 1.1508. ---] The sub-DLA at $z=1.1508$ is a very complex
system, spanning $800$~{\kms} for absorption in many low ionization
transitions.  It has many molecular hydrogen lines detected.  Our
identifications for the molecules follow exactly those of
\citet{rei03}.  The saturated region of the sub-DLA profile spans
$\sim 1100$~{\kms}.  A separate component at $\sim -900$~{\kms} is
likely also to be {\Lya}, but with no associated metals in the {\it
HST}/STIS spectrum.  Many neutral species of \ion{C}{1} and \ion{N}{1}
are detected in a single, narrow component at $\sim 0$~{\kms},
presumably associated with the bulk of the \ion{H}{1}.  Features at
other velocities seen in the neutral transitions panels in Figure~20
are not related to this system (see line identifications in
Table~\ref{tab:tab3}).  \ion{O}{1}~$1302$ is blended with {\Lya} at
$z=1.3039$, and cannot be measured, and no other strong \ion{O}{1}
transitions are covered.  Many singly ionized transitions are detected
for the system, both in the $~0$~{\kms} component and at many other
velocities to the blue.  These match the \ion{Mg}{2} and \ion{Fe}{2}
from the VLT/UVES spectrum \citep{lev03}.  Strong \ion{Si}{3}
absorption is detected in many components, spanning the full velocity
range.  {\SiIVdblt} is detected as well, but there appears to be
significant blending with \ion{Si}{4}~$\lambda 1394$, because
\ion{Si}{4}~$\lambda 1403$ is much weaker at some velocities.
{\CIVdblt} is also detected in the VLT/UVES spectrum.  {\NVdblt} is
covered, but not detected, and {\OVIdblt} is not covered.

\item[\it z = 1.3849. ---] An \ion{O}{6} doublet detected in the {\it
HST}/STIS spectrum at $z=1.3849$ has its $\lambda 1038$ member blended
to the blue with {\Lyb} at $1.4124$.  {\CIVdblt} is detected at the
same redshift in the VLT/UVES spectrum \citep{lev03}.  The
corresponding, relatively weak (unsaturated) {\Lya} line lies blueward
of a larger saturated {\Lya} feature.  \ion{C}{3} may also be
detected.

\item[\it z = 1.4163. ---] There is a possible \ion{O}{6} system at
$z=1.4163$, as reported by \citet{lev03}, however neither member of
the \ion{O}{6} doublet aligns with the {\Lya} and {\Lyb}, which are
also detected in the STIS spectrum.  There could be \ion{C}{3}
associated with this system, but it is badly blended with a probable
{\Lya} line, thus its alignment cannot be verified.  Therefore, there
is not convincing evidence for detected metals in this system.

\item[\it z = 1.6020. ---] The \ion{O}{6} system at $z=1.6020$ is well
aligned and is detected in {\Lyb} and {\Lyg} in the {\it HST}/STIS
spectrum.  {\Lya}, and possibly \ion{C}{4}, is detected in the
VLT/UVES spectrum \citep{lev03}.

\item[\it z = 1.6668. ---] There is a possible \ion{O}{6} system at
$z=1.6668$, though its identity is somewhat uncertain.
\ion{O}{6}~$\lambda 1038$ is blended with a possible {\Lya} line, such
that the profiles of the doublet members cannot be compared. {\Lyb} is
not detected at the position, however {\Lya} is detected, with about
the same profile shape and equivalent width as the \ion{O}{6}.  Thus,
this could be a fairly typical associated \ion{O}{6} absorber.

\item[\it z = 1.6736. ---] There is an \ion{O}{6} system at $z=1.6736$
detected in the {\it HST}/STIS spectrum.  \ion{O}{6}~$\lambda 1038$ is
blended to the red with a possible {\Lya} line.  {\Lyb} is detected,
as are higher order Lyman series lines, but the latter are blended
with other transitions (see Table~\ref{tab:tab3}).  {\CIVdblt} and
{\Lya} are detected in the VLT/UVES spectrum \citep{lev03}.

\item[\it z = 1.6971. ---] The associated \ion{O}{6} absorber at
$z=1.6971$ also has \ion{C}{3}~$\lambda 977$ and probably
\ion{S}{6}~$\lambda 933$ detected.  \citet{lev03} also detected
{\Lya}, {\SiIVdblt}, {\CIVdblt}, and {\NVdblt} in their VLT/UVES
spectrum.  The {\Lya} profile is not black, again a signature of
associated \ion{O}{6} absorbers.

\item[\it z = 1.7358. ---] A possible \ion{O}{6} system at $z=1.7358$
cannot be confirmed by other features in the {\it HST}/STIS spectrum.
Although {\Lyb} is covered, it is blended with Galactic
\ion{Mg}{2}~$\lambda 2803$.  If real, this system would be
significantly redshifted relative to the emission redshift of the
quasar.

\end{description}

\subsection{PG1206+459 ($z_{em}=1.16254$)\label{sec:pg1206}}

This quasar was observed by {\it HST}/FOS as part of the Quasar
Absorption Line Key Project, but they had difficulty in identifying
many of the lines due to uncertain continuum fitting \citep{jan98}.
They found evidence for an extensive metal-line system separated into
subsystems at $z=0.9254$, 0.9277 and 0.9342, even at low resolution.
\citet{chu99} modeled this system based on the {\it HST}/FOS data,
along with Keck/HIRES data \citep{chu01}.  \citet{din03a} refined
these models once the {\it HST}/STIS spectra were available.

\citet{chu99} observed this quasar with {\it HST}/STIS to facilitate a
detailed study of the systems at $z\sim0.927$.  The {\it HST}/STIS
spectrum covers the {\Lya} forest blueward of $2628$~{\AA}, but does
not cover any higher order Lyman series lines.  There are only three
metal-line systems found in the {\it HST}/STIS spectrum: the extensive
metal-line system at $z\sim0.927$ mentioned above, a \ion{C}{4}
system, and an associated \ion{N}{5} system.  Figure~28--32 presents
system plots for these metal-line systems, detected toward
PG$1206+459$.

\begin{description}

\item[\it z = 0.7338. ---] There is a weak \ion{C}{4} doublet at
$z=0.7338$.  The \ion{C}{4}~$\lambda 1548$ is blended with
\ion{Si}{4}~$\lambda 1394$ from the subsystem at $z=0.9254$, and also
to the red with an unknown blend, probably {\Lya}.  The large feature
blueward of the \ion{C}{4}~$\lambda 1551$ transition is
\ion{Si}{4}~$\lambda 1394$ at $z=0.9276$.  \ion{Mg}{2} is not detected
in the Keck/HIRES spectrum, nor any other low ionization transitions
in the {\it HST}/STIS spectrum.  However, the system is confirmed by
{\Lya} in the {\it HST}/FOS spectrum.

\item[\it z = 0.9254. ---] The complex metal-line system at $z=0.9277$
is spread over more than $1000$~{\kms}.  \citet{din03a} separated it
into three systems at $z=0.9254$ (System A), $z=0.9277$ (System B),
and $z=0.9342$ (System C).  They suggest that these systems are
produced by three different galaxies in a group.  System A, at
$z=0.9254$, has detected \ion{C}{2}, \ion{Si}{3}, {\SiIVdblt},
{\CIVdblt}, and {\NVdblt} in the {\it HST}/STIS spectrum.  Many other
transitions, including {\OVIdblt} and Lyman series lines, were
detected in the {\it HST}/FOS spectrum.  \ion{Si}{4}~$\lambda 1394$ is
blended with \ion{C}{4}~$\lambda 1548$ from the $z=0.7338$ \ion{C}{4}
system.  {\Lya} is saturated and covers the entire velocity range
between this system and System B.

\item[\it z = 0.9276. ---] System B, at $z=0.9276$, produces a partial
Lyman limit break in the {\it HST}/FOS spectrum.  It is detected in
several transitions of \ion{Si}{2}, \ion{C}{2}, \ion{Si}{3},
{\SiIVdblt}, {\CIVdblt}, and {\NVdblt}.  Galactic \ion{Fe}{2}~$\lambda
2344$ is blended with the red wing of the {\Lya} profile.
\ion{C}{2}~$\lambda 1335$ is blended to the red with an unidentified
line, probably {\Lya}.  The feature redward of \ion{Si}{4}~$\lambda
1394$ is \ion{C}{4}~$\lambda 1551$ from the $z=0.7338$ system.
\ion{C}{4} is self-blended, with the \ion{C}{4}~$\lambda 1551$ from
System A superimposed on the \ion{C}{4}~$\lambda 1548$ from System B.

\item[\it z = 0.9343. ---] System C, at $z=0.9343$ is unusual in its
high velocity relative to the main system, System B.  For this system,
which would be considered a single-cloud, weak \ion{Mg}{2} absorber if
isolated from the other systems, is detected in {\Lya}, \ion{Si}{2},
\ion{C}{2}, \ion{Si}{3}, {\SiIVdblt}, and {\CIVdblt}.
\ion{N}{5}~$\lambda 1239$ is blended with \ion{N}{5}~$\lambda 1243$
from System B, but \ion{N}{5}~$\lambda 1243$ is not detected in a
clean part of the spectrum.  Again, \ion{O}{6} and various other
transitions are detected in the {\it HST}/FOS spectrum.

\item[\it z = 1.0281. ---] The \ion{N}{5} system at $z=1.0281$ is
almost certainly intrinsic judging by its broad profile and the
``non-black'' {\Lya} line, evidence for partial covering.  No other
metals are detected besides the \ion{N}{5}.

\end{description}

\subsection{PG1241+176 ($z_{em}=1.273$)\label{sec:pg1241}}

This was also a quasar studied for the Quasar Absorption Line Key
Project \citep{jan98}, and a Keck/HIRES spectrum was obtained by
\citet{chu01}.  Particular \ion{Mg}{2} systems, with optical and UV
coverage, were studied in \citet{archive1}.  The $z=0.5505$, 0.5584
and 0.8954 systems were modeled in detail by \citet{din05}.

This quasar was observed with {\it HST}/STIS by Churchill et al. in
order to provide constraints on the physical conditions in \ion{Mg}{2}
absorbers.  {\Lya} forest lines appear in {\it HST}/STIS spectrum
quasar spectrum blueward of 2763~{\AA}, and the {\Lyb} forest is
covered blueward of 2332~{\AA}.  We find evidence for six metal-line
systems: one strong \ion{Mg}{2} absorber, two weak \ion{Mg}{2}
absorbers, one weak \ion{C}{4}, and two associated \ion{O}{6}
absorbers.  The system plots for these metal-line systems are shown in
Figures~33--38.

\begin{description}

\item[\it z = 0.5507. ---] \citet{jan98} found a metal-line system at
$z=0.5507$ with strong \ion{C}{4} detected.  This system corresponds
to a strong \ion{Mg}{2} absorber detected in a Keck/HIRES spectrum
\citep{chu01}, which also has detected \ion{Ca}{2}, \ion{Fe}{2}, and
\ion{Mg}{1}.  In the {\it HST}/STIS spectrum, we also find
\ion{Fe}{2}~$\lambda 1608$, \ion{Al}{2}~$\lambda 1671$, and
{\CIVdblt}.  There is a $3\sigma$ detection of \ion{Al}{2}~$\lambda
1671$ at $v\sim150$~{\kms}, which coincides with a satellite component
of {\MgIIdblt} in the Keck/HIRES spectrum \citep{din05}.

\item[\it z = 0.5584. ---] For the multiple-cloud, weak \ion{Mg}{2}
absorber at $z=0.5584$, {\CIVdblt} was detected in the {\it HST}/STIS
spectrum, but no low ionization transitions were detected, though
\ion{Al}{2}~$\lambda 1671$, {\AlIIIdblt}, and \ion{Si}{2}~$\lambda
1527$ were covered.

\item[\it z = 0.7577. ---] A possible weak \ion{C}{4} doublet at
$z=0.7577$ cannot be confirmed, since {\Lya} is not covered in the
{\it HST}/STIS or in the previous {\it HST}/FOS spectrum.

\item[\it z = 0.8954. ---] The single-cloud, weak \ion{Mg}{2} absorber
at $z=0.8954$ has detected {\Lya}, \ion{Si}{3}~$\lambda 1207$,
\ion{Si}{4}~$\lambda 1394$, and {\CIVdblt}.  The \ion{Si}{4}~$\lambda
1403$ is affected by a blend.  Although there is a detection at the
position of \ion{Si}{2}~$\lambda 1260$, \citet{din05} note that this
may be a blend, since it is difficult to explain the strength of this
feature in comparison to the weaker {\MgIIdblt} lines.

\item[\it z = 1.2152. ---] The \ion{O}{6} system, at $z=1.2152$, found
by \citet{jan98} is confirmed here, showing {\OVIdblt} and {\Lya} in
the {\it HST}/STIS spectrum.  Although a feature is detected at the
position of \ion{N}{5}~$\lambda 1239$, also noted by \citet{jan98},
this is inconsistent with the lack of detection of \ion{N}{5}~$\lambda
1243$, and so cannot be confirmed.

\item[\it z = 1.2717. ---] \citet{jan98} found a strong associated
\ion{O}{6} absorber at $z=1.2717$ in their {\it HST}/FOS spectrum, for
which {\Lya} and {\Lyb} were also detected.  In the higher resolution
{\it HST}/STIS spectrum, we confirm these detections, but do not find
any other associated absorption.  The relatively weak {\Lya}, as
compared to {\OVIdblt} is characteristic of a class of associated
\ion{O}{6} absorbers \citep{gan07}.

\end{description}

\subsection{PG~1248+401 ($z_{em}=1.030$)\label{sec:pg1248}}

This quasar was studied by the {\it HST} Quasar Absorption Line Key
Project using G190H and G270H spectra obtained with FOS \citep{jan98},
who found metal-line systems at $z=0.3946$, 0.7660, 0.7732 and 0.8553.
The latter two systems correspond to strong \ion{Mg}{2} absorbers,
modeled by \citet{din05} based on Keck/HIRES data \citep{chu01} and
the {\it HST}/STIS spectrum presented here.  Another system, at
$z=0.7011$, with only {\CIVdblt} and {\Lya} detected, was discussed in
\citet{mil06}.

\citet{chu99} obtained a spectrum of this quasar with {\it HST}/STIS
in order to provide constraints on the physical conditions in the two
strong \ion{Mg}{2} absorbers along the line of sight.  In the {\it
HST}/STIS spectrum, the {\Lya} forest is only covered up to a
wavelength of $2468$~{\AA}, and higher order Lyman series lines are
not covered.  We find evidence for two strong \ion{Mg}{2} and two weak
\ion{C}{4} systems in the {\it HST}/STIS spectra.  These systems are
shown in Figures~39--45.

\begin{description}

\item[\it z = 0.3946. ---] \ion{Al}{2}~$\lambda 1671$ and
\ion{Al}{3}~$\lambda$1855 are the only prominent transitions from the
possible $z=0.3946$ system that are covered in the {\it HST}/STIS
spectrum.  Although there is a detection at the position of
\ion{Al}{2}~$\lambda 1671$, the lack of detected {\AlIIIdblt} and the
lack of detection of {\CIVdblt} in the FOS spectrum leads us to
question the reality of this system.  Thus we prefer an identification
of the feature at $2330.09$~{\AA} as {\Lya}.

\item[\it z = 0.5648. ---] A \ion{C}{4} doublet is found at $z=0.5648$
in the {\it HST}/STIS spectrum.  It is confirmed by a {\Lya} detection
in the G190H {\it HST}/FOS spectrum \citep{jan98}.  The weak feature
at the expected position of \ion{Si}{2}~$\lambda 1527$ is unlikely to
be \ion{Si}{2} because of the absence of the stronger
\ion{Si}{2}~$\lambda 1260$ transition in the FOS spectrum.

\item[\it z = 0.6174. ---] There is a very weak possible \ion{C}{4}
doublet at $z=0.6174$ in the {\it HST}/STIS spectrum, however the
alignment is not perfect.  Also, there is no {\Lya} detected in the
FOS spectrum to a rest frame equivalent width limit of $0.11$~{\AA}.

\item[\it z = 0.7011. ---] The \ion{C}{4} absorption system at
$z=0.7011$ is confirmed by a {\Lya} detection in the FOS spectrum, but
there are no additional detections in the {\it HST}/STIS spectrum.

\item[\it z = 0.7728. ---] The strong \ion{Mg}{2} system at $z=0.7728$
has detected \ion{Al}{2}~$\lambda 1671$, \ion{Si}{2}~$\lambda 1304$
and \ion{Si}{2}$\lambda 1527$, \ion{C}{2}~$\lambda 1335$, {\SiIVdblt},
and {\CIVdblt} in the STIS E230M spectrum.  \ion{O}{1}~$\lambda 1302$
may be detected in the strongest low ionization component, but the
spectrum is very noisy in that region.

\item[\it z = 0.7760. ---] We cannot confirm a system at $z=0.7760$,
which was suggested by \citet{jan98}, however we note that we also do
not detect {\CIVdblt} to a $3\sigma$ rest frame equivalent width limit
of $0.02$~{\AA}.  It is still possible that this system exists, but is
collisionally ionized, as proposed by \citet{jan98}.

\item[\it z = 0.8545. ---] The strong \ion{Mg}{2} system at $z=0.8545$
has detected \ion{Si}{2}~$\lambda 1260$, \ion{C}{2}~$\lambda 1335$,
{\SiIVdblt}, {\CIVdblt}, and {\NVdblt}.  Kinematically, this system is
interesting, with satellite components in low ionization gas (offset
$\sim 300$~{\kms} from the main absorption) having corresponding,
broader \ion{C}{4} absorption.

\end{description}

\subsection{CSO~$873$ ($z_{em}=1.022$)\label{sec:cso873}}

This quasar was previously studied through G190H and G270H {\it
HST}/FOS spectra \citep{bah96}, but only three metal-line systems were
indicated, at $z=0.2891$, 0.6606 and 1.0022.  The $z=0.6606$ coincides
with a strong \ion{Mg}{2} absorber \citep{chu01}, and it was modeled
by \citet{din05}.

The {\it HST}/STIS spectrum was obtained in a program (P.I. Churchill)
to facilitate a study of the $z=0.6606$ \ion{Mg}{2} absorber.  This
spectrum covers the {\Lya} forest up to a wavelength of $2458$~{\AA},
and does not cover any {\Lyb} lines.

\begin{description}

\item[\it z = 0.6611. ---] Besides Galactic absorption, in the {\it
HST}/STIS E230M spectrum, we detect metal-line absorption from only
the system at $z=0.6611$, which corresponds to a full Lyman limit
break seen in a {\it HST}/FOS G160L spectrum \citep{bah93,bah96}.  A
system plot is shown in Figure~46.  From this system we find
\ion{Al}{2}~$\lambda 1671$, \ion{Si}{2}~$\lambda 1527$, and
{\CIVdblt}.  \ion{Si}{4} is not detected.  In fact, the \ion{C}{4}
absorption at the velocity of the low ionization absorption is
relatively weak, and this system classifies as
``\ion{C}{4}--deficient'' \citep{din05}.  The strongest \ion{C}{4}
absorption is $\sim 200$~{\kms} redward of the strongest low
ionization component.  \citet{jan98} note a complex of {\Lya} lines
ranging from $z=0.6692$ to $0.6771$.  They found no metals associated
with this complex in the low resolution {\it HST}/FOS spectrum.  We
note that with the STIS data, it is possible to set a stronger limit,
with a rest frame $3\sigma$ equivalent width limit $W_r(1548) <
0.02A$.

\end{description}

\subsection{PG$1634+706$ ($z_{em}=1.334$)\label{sec:pg1634}}

This was one of the {\it HST} QSO Absorption Line Key Project quasars,
and its {\it HST}/FOS G270M spectrum was presented in \citet{bah96}.
In that spectrum, three extensive metal-line systems are present at
$z=0.9060$, 0.9908 and 1.0417.  Also, there are possible \ion{C}{4}
doublets at $z=0.5582$, 0.6540, 0.6790 and 0.7796, and a possible
\ion{O}{6} doublet at $z=1.3413$.  In a Keck/HIRES optical spectrum,
strong \ion{Mg}{2} absorption is detected from the $z=0.9908$ system,
and weak \ion{Mg}{2} absorption from the $z=0.6540$, 0.9060 and 1.0417
systems \citep{chu01,chu99,cha03}.  Weak \ion{Mg}{2} is also detected
at $z=0.8181$ in the optical spectrum \citep{chu99}.  The three
single-cloud, weak \ion{Mg}{2} absorbers at $z=0.8181$, 0.9056 and
0.6534 were modeled by \citet{cha03}, the multiple-cloud, weak
\ion{Mg}{2} absorber at $z=1.0414$ by \citet{zon04}, and the strong
\ion{Mg}{2} absorber at $z=0.9902$ by \citet{din03b}.

Two programs contribute different wavelength coverage of this quasar
with {\it HST}/STIS.  Burles (1997, HST Proposal ID \#7292) observed
the quasar in order to measure the primordial D/H ratio.
\citet{jan98} provided redward coverage in order to conduct a detailed
study of the {\Lya} forest.  In the {\it HST}/STIS spectra, {\Lya} is
covered up to $2837$~{\AA}, {\Lyb} up to $2395$~{\AA}, and {\Lyg} in
the small region up to $2268$~{\AA}.  There is detected absorption
from the strong \ion{Mg}{2} absorber, and from the four weak
\ion{Mg}{2} absorbers mentioned above.  There are also several
possible \ion{C}{4} systems and a possible \ion{O}{6} system, as well
as an interesting ``\ion{C}{3}/\ion{Si}{3}-only'' system.  There is a
possible intrinsic \ion{N}{5} system and an associated \ion{O}{6}
system observed toward this quasar as well.  We report on these
systems here, and shown system plots in Figures~47--56.

\begin{description}

\item[\it z = 0.2854. ---] We can only tentatively claim a \ion{C}{4}
doublet at $z=0.2854$.  The \ion{C}{4}~$\lambda 1548$ and
\ion{C}{4}~$\lambda 1551$ profiles match reasonably well, but
\ion{C}{4}$\lambda 1551$ may be a bit too strong at some velocities,
and there is no coverage of transitions that could confirm this
identification.  This could relate to a blend with the
\ion{C}{3}~$\lambda 977$ from the $z=1.0415$ system (see below).

\item[\it z = 0.6513. ---] There is also a probable, narrow \ion{C}{4}
doublet at $z=0.6513$.  The \ion{C}{4}~$\lambda 1550$ transition is
blended with \ion{C}{4}~$\lambda 1548$ from the $z=0.6535$ system.  If
this is real, it is unusual in the weakness of {\Lya} relative to
\ion{C}{4}.  A weak, narrow \ion{N}{5} doublet may also be detected.

\item[\it z = 0.6535. ---] The {\it HST}/STIS spectrum shows weak
absorption in several low and intermediate transitions related to the
$z=0.6535$ single-cloud, weak \ion{Mg}{2} absorber.  The \ion{C}{4}
has two components redward of the dominant low ionization absorption,
and \ion{Si}{4} is weaker but has similar kinematics.

\item[\it z = 0.8182. ---] The single-cloud, weak \ion{Mg}{2} absorber
at $z=0.8182$ has detected \ion{Si}{2}, \ion{C}{2}, \ion{Si}{4}, and
\ion{C}{4} absorption, all centered at the same velocity as the
\ion{Mg}{2}.  \ion{Si}{3}~$\lambda 1206$ is blended with {\Lyb} from a
system at $z=1.1382$, and is apparent as a small depression in the red
wing of that feature.

\item[\it z = 0.9056. ---] For another single-cloud, weak \ion{Mg}{2}
absorber at $z=0.9056$, {\Lya}, {\Lyb}, \ion{C}{2}, \ion{Si}{2},
\ion{Si}{4}, \ion{C}{4}, \ion{N}{5}, and \ion{O}{6} are detected in
the {\it HST}/STIS spectrum.  The coverage of \ion{N}{5} and
\ion{O}{6} is unusual for single-cloud, weak \ion{Mg}{2} absorbers and
is important for considerations of the phase structure of these
objects.

\item[\it z = 0.9645. ---] There is an interesting system detected in
the {\it HST}/STIS spectrum at $z=0.9645$.  Most of the common metal
transitions have good coverage, but only \ion{C}{3} and \ion{Si}{3}
are detected.  {\Lya} and {\Lyb} confirm the reality of this system.

\item[\it z = 0.9904. ---] The \ion{C}{4}-deficient, strong
\ion{Mg}{2} absorber at $z=0.9904$ has detections in the {\it
HST}/STIS spectrum of the first four Lyman series lines, \ion{O}{1},
and a variety of singly to triply ionized transitions, including
relatively weak \ion{C}{4} absorption.  \ion{N}{5} and \ion{O}{6} are
covered, but not detected.  The {\Lyd} line is detected, but suffers
from a blend to the blue with \ion{H}{1}~$\lambda 926$ from the system
at $z=1.0415$.  This system is of interest because the higher
ionization transitions appear redward of the lower ionization
transitions, indicating a density gradient across the system
\citep{din05}.

\item[\it z = 1.0415. ---] The $z=1.0415$ multiple-cloud, weak
\ion{Mg}{2} system has a partial Lyman limit break detected in a {\it
HST}/STIS G230M spectrum.  The higher order Lyman series lines are
detected down to the break in the {\it HST}/STIS E230M spectrum,
though past \ion{H}{1}~$\lambda 919$ they suffer badly from blending
with the Lyman series lines from the $z=0.9904$ system.  Metal lines
are detected in \ion{Si}{2}, \ion{C}{2}, \ion{Si}{3}, \ion{Si}{4}, and
\ion{O}{6} from each of two subsystems.  The low ionization
transitions are centered at different velocities than the high
ionization transitions.  \ion{C}{3}~$\lambda 977$ is also detected,
but may be blended with the possible \ion{C}{4} doublet at $z=0.2854$.
\ion{C}{4} is detected in a low resolution {\it HST}/FOS spectrum, but
is redward of the {\it HST}/STIS coverage.

\item[\it z = 1.1408. ---] There is a possible \ion{O}{6} absorber at
$z=1.1408$, however its \ion{O}{6}~$\lambda 1032$ transition is too
strong relative to the \ion{O}{6}~$\lambda 1038$ transition.  Although
{\Lya}, {\Lyb}, and {\Lyg} are detected at this velocity, their
relative strengths are also inconsistent, perhaps due to unknown
blends.

\item[\it z = 1.3413. ---] An \ion{O}{6} doublet is detected from a
system at $z=1.3413$, which is slightly redward of $z_{em}$.  It is
confirmed by a very weak {\Lya} detection, but {\Lyb} is badly blended
and provides no additional constraint.  This is another example of an
associated \ion{O}{6} system (e.g. \citet{gan07}).

\end{description}

\subsection{PG$1718+481$ ($z_{em}=1.084$)\label{sec:pg1718}}

The E230M observation of this quasar with {\it HST}/STIS was proposed
by Burles et al. (1997, HST Proposal ID \#7292) to facilitate a study
of the deuterium to hydrogen ratio in a system at $z=0.701$
\citep{kir01}.  An {\it HST}/STIS G230M spectrum was also obtained,
covering the Lyman series for this system \citep{kir01}.  Earlier,
this quasar was observed as part of the HST Quasar Absorption Line Key
Project.  \citet{jan98} found metal-line systems at $z=0.8929$, 1.0323
and 1.0872.

In the {\it HST}/STIS spectrum, {\Lya} is covered up to $2535$~{\AA},
{\Lyb} up to $2138$~{\AA}, and {\Lyg} up to $2036$~{\AA}.  All Lyman
series lines are covered below $1902$~{\AA}.  In addition to the
$z=0.701$ system, for which this observation was obtained, we find
evidence for four definite and two possible \ion{O}{6} absorbers.
These systems are all shown in Figures~57--64.

\begin{description}

\item[\it z = 0.7011. ---] The $z=0.7011$ Lyman limit system has
{\Lya} and \ion{Si}{3} detected in the {\it HST}/STIS spectrum.  Weak
\ion{C}{2}~$\lambda 1335$ also appear to be detected, consistent with
the weak \ion{Mg}{2} absorption that \citet{kir01} found in a
HIRES/Keck spectrum.  {\CIVdblt} and {\SiIVdblt} are covered, but not
detected.

\item[\it z = 0.8928. ---] An \ion{O}{6} absorber is found at
$z=0.8928$ in the STIS spectrum, accompanied by detections of
\ion{Si}{3}, {\Lya}, and {\Lyb}.  \citet{jan98} also found \ion{C}{3}
in a low resolution {\it HST}/FOS spectrum.

\item[\it z = 1.0065. ---] Another \ion{O}{6} system is detected at
$z=1.0065$ in the {\it HST}/STIS spectrum.  {\Lya} and {\Lyb} are also
detected, confirming this \ion{O}{6} doublet.

\item[\it z = 1.0318. ---] A stronger \ion{O}{6} absorber is detected
at $z=1.0318$, corresponding to the system found by \citet{jan98}.  In
the {\it HST}/STIS spectrum the first five Lyman series lines are
detected, along with \ion{C}{3}, {\NVdblt}, and {\OVIdblt}.

\item[\it z = 1.0507. ---] There is a feature detected (just above
$5\sigma$) at the position of \ion{O}{6}~$\lambda 1032$ for a system
at $z=1.0507$ with detected {\Lya} and {\Lyb} absorption.  The
\ion{O}{6}~$\lambda 1038$ is not detected, even at $2\sigma$, so it is
uncertain whether there are detected metals for this system.

\item[\it z = 1.0548. ---] The $z=1.0548$ system, with {\Lya}, {\Lyb},
{\Lyg}, and {\Lyd} detected in the {\it HST}/STIS spectrum, appears to
be an \ion{O}{6} absorber, with weak \ion{C}{3} absorption.  The
\ion{O}{6} doublet cannot be confirmed because the \ion{O}{6}~$\lambda
1032$ member is blended with {\Lyb} at $z=1.0674$.

\item[\it z = 1.0867. ---] There is an associated system with
$z=1.0867$, detected in {\Lya}, {\Lyb}, and {\Lyg}.  The {\Lya}
profile has a black, saturated region at this redshift. The $z=1.0867$
component has detections in \ion{C}{3} and \ion{Si}{3}.

\item[\it z = 1.0874. ---] There is also a non-black component at
$z=1.0874$.  This system detected {\OVIdblt}, and resembles many other
associated \ion{O}{6} systems.

\end{description}

\section{Summary of Systems}\label{sec:summary}

This paper presents a catalog that is useful for investigations of
metal-line systems and the {\Lya} forest in eight of the highest
S/N-ratio E230M spectra from the {\it HST}/STIS archive.  We have
identified a total of 56 metal-line systems (with at least one metal
line detected) toward the eight quasars listed in
Table~\ref{tab:tab1}, as presented in Figures~9--64.  For key
transitions, Table~\ref{tab:tab4} tabulates in how many of the 56
systems a given transition was covered and in how many it was
detected.  We must caution that clearly these numbers apply only to
this particular dataset and are highly dependent on the detection
limits of the spectra, to the exact wavelength coverage, and to biases
toward certain kinds of quasars for the sample.  In this sample of
eight quasars, there may be a preference toward strong low ionization
absorbers, since many of them were chosen for a program to study the
\ion{C}{4} associated with \ion{Mg}{2} absorption (see
Table~\ref{tab:tab1}).

Nonetheless, it is useful to note from Table~\ref{tab:tab4} that for
about half of the 56 systems we have coverage of the \ion{C}{4} and
\ion{O}{6} doublets, and that in those cases these high ionization
transitions were detected $\sim 90$\% of the time.  Intermediate
ionization transitions such as \ion{C}{3}~$\lambda$977 and
\ion{Si}{3}~$\lambda$1207 were also covered about half the time, and
were detected in 50\%/58\% of the sight lines,
respectively. \ion{C}{2}~$\lambda$1335 is also covered for about half
of the systems, and was detected 56\% of the time, while
\ion{Si}{2}~$\lambda$1260 was only detected in 36\% of the sightlines
for which it was covered.  We conclude that almost all 1 $<$ $z$ $<$
1.7 metal-line systems have high ionization absorption detected, but
that only half have detected low ionization absorption.  These numbers
are consistent with the findings of \citet{mil06}, that half of all
\ion{C}{4} absorbers at low redshift have detected low ionization
absorption (sometimes weak), and that the other half do not.  This is
not surprising since the samples used for these studies have
considerable overlap.

Studies of the statistics of the {\Lya} forest can be significantly
affected by metal lines in the forest.  With our catalog we are able
to give an indication of the level of contamination by metal lines in
the $1.0 < z < 1.7$ forest.  Specifically, we scanned the wavelength
regions of our catalog that are in the {\Lya} forest of the quasar,
but are redward of the {\Lyb} forest.  Table~\ref{tab:tab1} lists how
many {\Lya} lines and how many metal lines were detected in this
region for each quasar.  It also shows that 9-10 of the features
detected in the E230M quasar spectra are actually Galactic absorption.
We conclude that of the $1.0 < z < 1.7$ {\Lya} forest lines detected
toward each of the quasars (redward of the {\Lyb} forest), 20-33\% of
the lines in the {\Lya} forest are actually metal lines.  This
"contamination" is significant enough that it is important to remove
it before computing the statistics of the {\Lya} lines, and is clearly
itself a complicated function of redshift.

\section*{Acknowledgments}

We acknowledge support from NASA under grant NAG5-6399 NNG04GE73G and
by the National Science Foundation under grant AST
04-07138. J. R. M. and R. S. L. were partially funded by the NSF
Research Experience for Undergraduates (REU) program.  DT and DK were
supported in part by STScI grants HST-AR-10688 and HST-AR-10288.  SB
was supported by the NSF/REU program in the summer of 2005.

\label{lastpage}
\clearpage


\begin{table*}
 \centering
 \caption{{\it HST}/STIS E230M Quasar Observations}\label{tab:tab1}
 \begin{tabular}{@{}llcclcccccc@{}}
 \hline
 \hline
 QSO ID & $z_{em}$ & \multicolumn{2}{c}{$S/N^{a}$} & PI & Proposal ID & $\lambda_{lo}$ & $\lambda_{up}$ & $N_{tot}$$^b$ & $N_{metal}$$^c$ & $N_{Gal}$$^d$ \\
        &          &   2382~\AA  &   2796~\AA      &    &             &    (\AA)       &    (\AA)       &               &                 &               \\
 \hline
 PG 0117+210  & 1.491 &  8.5 & 14.2 & Jannuzi        & 8673      & 2278.7 & 3117.0 & 104 & 31 & 10 \\
 HE 0515-4414 & 1.713 &  8.1 & 22.0 & Reimers        & 8288      & 2274.8 & 3118.9 &  63 & 13 &  9 \\
 PG 1206+459  & 1.160 &  8.4 & 15.4 & Churchill      & 8672      & 2277.1 & 3116.4 &  85 & 33 &  9 \\
 PG 1241+176  & 1.273 &  5.2 & 20.0 & Churchill      & 8672      & 2276.9 & 3066.0 &  62 & 20 & 10 \\
 PG 1248+401  & 1.030 &  8.1 & 9.6  & Churchill      & 8672      & 2276.7 & 3107.7 &  41 & 14 & 10 \\
 CSO 873$^e$  & 1.022 &  8.0 & 10.4 & Churchill      & 8672      & 2278.4 & 3119.2 &  26 &  3 &  8 \\
 PG 1634+706  & 1.334 & 19.8 & 28.5 & Jannuzi/Burles & 8312/7292 & 1860.7 & 3117.0 & 121 & 33 & 18 \\
 PG 1718+481  & 1.083 & 15.9 & ...  & Burles         & 7292      & 1848.4 & 2672.7 &  81 & 15 & 14 \\
 \hline
 \end{tabular}
 \\
 $^a$ $S/N$ per pixel at 2382~\AA\ and 2796~\AA,
 $^b$ Total number of absorption features between \Lya\ and \Lyb\ emission lines of quasar,
 $^c$ Number of identified metal absorption features between \Lya\ and \Lyb\ emission lines of quasar,
 $^d$ Number of Galactic metal absorption features in entire observed wavelength region,
 $^e$ Coordinate of this quasar: RA: 13:19:56.3 Dec: +27:28:09 (J2000.0).
\end{table*}

\clearpage

\begin{table*}
 \centering
 \caption{Detected Transitions}\label{tab:tab2}
 \begin{tabular}{ccc}
 \hline
 \hline
 Transition                 & $\lambda_{rest}$$^a$ &  Oscillator Strength$^a$ \\
                            & (\AA)                &                          \\
 \hline
 \Lya                       &  1215.6701       &  0.416400  \\
 \ion{H}{1}$\lambda$1026    &  1025.7223       &  0.079120  \\
 \ion{H}{1}$\lambda$972     &   972.5368       &  0.029000  \\
 \ion{H}{1}$\lambda$950     &   949.7431       &  0.013940  \\
 \ion{H}{1}$\lambda$938     &   937.8035       &  0.007799  \\
 \ion{H}{1}$\lambda$931     &   930.7483       &  0.004814  \\
 \ion{H}{1}$\lambda$926     &   926.2257       &  0.003183  \\
 \ion{H}{1}$\lambda$923     &   923.1504       &  0.002216  \\
 \ion{H}{1}$\lambda$921     &   920.9631       &  0.001605  \\
 \ion{H}{1}$\lambda$919     &   919.3514       &  0.00120   \\
 \ion{C}{1}$\lambda$1158    &  1157.910        &  0.021780  \\
 \ion{C}{1}$\lambda$1189    &  1188.833        &  0.016760  \\
 \ion{C}{1}$\lambda$1277    &  1277.245        &  0.096650  \\
 \ion{C}{1}$\lambda$1329    &  1328.833        &  0.058040  \\
 \ion{C}{2}$\lambda$1036    &  1036.3367       &  0.1231    \\
 \ion{C}{2}$\lambda$1335    &  1335.31         &  0.127     \\
 \ion{C}{2}$*\lambda$1336b  &  1335.6627       &  0.01277   \\
 \ion{C}{3}$\lambda$977     &   977.020        &  0.7620    \\
 \ion{C}{4}$\lambda$1548    &  1548.195        &  0.190800  \\
 \ion{C}{4}$\lambda$1551    &  1550.770        &  0.095220  \\
 \ion{N}{1}$\lambda$1134b   &  1134.1653       &  0.01342   \\
 \ion{N}{1}$\lambda$1135    &  1134.9803       &  0.04023   \\
 \ion{N}{1}$\lambda$1201    &  1200.7098       &  0.04423   \\
 \ion{N}{2}$\lambda$1084    &  1083.990        &  0.1031    \\
 \ion{N}{5}$\lambda$1239    &  1238.821        &  0.157000  \\
 \ion{N}{5}$\lambda$1243    &  1242.804        &  0.078230  \\
 \ion{O}{6}$\lambda$1032    &  1031.927        &  0.132900  \\
 \ion{O}{6}$\lambda$1038    &  1037.616        &  0.066090  \\
 \ion{Mg}{1}$\lambda$2853   &  2852.964        &  1.810000  \\
 \ion{Mg}{2}$\lambda$2796   &  2796.352        &  0.6123    \\
 \ion{Mg}{2}$\lambda$2803   &  2803.531        &  0.3054    \\
 \ion{Al}{2}$\lambda$1671   &  1670.787        &  1.880000  \\
 \ion{Al}{3}$\lambda$1855   &  1854.716        &  0.539000  \\
 \ion{Si}{2}$\lambda$990    &   989.873        &  0.133000  \\
 \ion{Si}{2}$\lambda$1190   &  1190.416        &  0.250200  \\
 \ion{Si}{2}$\lambda$1193   &  1193.290        &  0.499100  \\
 \ion{Si}{2}$*\lambda$1195  &  1194.500        &  0.623300  \\
 \ion{Si}{2}$\lambda$1260   &  1260.422        &  1.007000  \\
 \ion{Si}{2}$\lambda$1304   &  1304.370        &  0.147300  \\
 \ion{Si}{2}$\lambda$1527   &  1526.707        &  0.11600   \\
 \ion{Si}{3}$\lambda$1207   &  1206.500        &  1.660000  \\
 \ion{Si}{4}$\lambda$1394   &  1393.755        &  0.528000  \\
 \ion{Si}{4}$\lambda$1403   &  1402.770        &  0.262000  \\
 \ion{P}{2}$\lambda$1153    &  1152.818        &  0.236100  \\
 \ion{S}{2}$\lambda$1251    &  1250.584        &  0.005453  \\
 \ion{Cr}{2}$\lambda$2056   &  2056.254        &  0.1403    \\
 \ion{Cr}{2}$\lambda$2062   &  2062.234        &  0.1049    \\
 \ion{Cr}{2}$\lambda$2066   &  2066.161        &  0.06982   \\
 \ion{Mn}{2}$\lambda$2577   &  2576.877        &  0.3508    \\
 \ion{Mn}{2}$\lambda$2594   &  2594.499        &  0.2710    \\
 \ion{Mn}{2}$\lambda$2606   &  2606.461        &  0.1927    \\
 \ion{Fe}{1}$\lambda$2524   &  2523.608        &  0.279000  \\
 \ion{Fe}{2}$\lambda$1608   &  1608.4449       &  0.055450  \\
 \ion{Fe}{2}$\lambda$2250   &  2249.8768       &  0.00182   \\
 \ion{Fe}{2}$\lambda$2261   &  2260.7805       &  0.00244   \\
 \ion{Fe}{2}$\lambda$2344   &  2344.214        &  0.109700  \\
 \ion{Fe}{2}$\lambda$2374   &  2374.4612       &  0.02818   \\
 \ion{Fe}{2}$\lambda$2383   &  2382.765        &  0.3006    \\
 \ion{Fe}{2}$\lambda$2587   &  2586.650        &  0.064570  \\
 \ion{Fe}{2}$\lambda$2600   &  2600.1729       &  0.22390   \\
 \ion{Zn}{2}$\lambda$2063   &  2062.664        &  0.252900  \\
 \hline
 \end{tabular}
\\
 $^a$ These numbers are from \citet{vvf96}.
\end{table*}

\clearpage

\begin{table*}
 \centering
 \caption{Sample -- Features with $5\sigma$ Identified in the {\it
          HST}/STIS E230M Spectra of 8 Quasars}\label{tab:tab3}
 \begin{tabular}{rllllcll}
 \hline
 \hline
 \# & $\lambda_{obs}$ & $W_{obs}$ & $\sigma(W_{obs})$ & $S$ & Line ID & z & Notes \\
 \hline
 \multicolumn{8}{c}{PG~$0117+213$ ($z_{em}=1.493$)} \\
 \hline
    1 & 2278.83 &     0.21 &     0.02 &     9.5 &     \ion{H}{1}$\lambda$972 &     1.3430 &            \\
    2 & 2285.29 &     0.97 &     0.04 &    25.7 &     \ion{C}{3}$\lambda$977 &     1.3390 &            \\
    3 & 2287.19 &     0.11 &     0.02 &     6.3 &                   {\Lya} ? &     0.8814 &            \\
    4 & 2288.62 &     1.62 &     0.03 &    49.4 &     \ion{C}{3}$\lambda$977 &     1.3430 & Probable blend with {\Lya} at $z$=0.8026 ? \\
    5 & 2292.90 &     0.14 &     0.02 &     7.9 &                   {\Lya} ? &     0.8861 &            \\
    6 & 2293.51 &     0.14 &     0.02 &     6.8 &                   {\Lya} ? &     0.8866 &            \\
    7 & 2295.47 &     0.47 &     0.02 &    21.4 &     \ion{H}{1}$\lambda$938 &     1.4478 &            \\
    8 & 2301.45 &     1.31 &     0.05 &    28.4 &                   {\Lya} ? &     0.8932 & Reddest component is \ion{H}{1}$\lambda$950 at $z$=1.4242 \\
    9 & 2306.53 &     0.05 &     0.01 &     4.2 &                   {\Lya} ? &     0.8973 &            \\
   10 & 2306.91 &     0.20 &     0.01 &    13.9 &    \ion{C}{2}$\lambda$1335 &     0.7290 &            \\
   11 & 2307.52 &     0.18 &     0.02 &    11.8 &    \ion{C}{2}$\lambda$1335 &     0.7290 &            \\
   12 & 2311.85 &     0.76 &     0.04 &    21.2 &    \ion{H}{1}$\lambda$1026 &     1.2543 &            \\
   13 & 2313.76 &     0.51 &     0.02 &    20.9 &                   {\Lya} ? &     0.9033 &            \\
   14 & 2316.57 &     0.09 &     0.01 &     6.0 &                   {\Lya} ? &     0.9056 &            \\
   15 & 2318.26 &     0.04 &     0.01 &     5.5 &                   {\Lya} ? &     0.9070 & Small contribution from \ion{N}{3}$\lambda$989 at $z$=1.3430 \\
   16 & 2319.01 &     0.38 &     0.02 &    17.5 &                   {\Lya} ? &     0.9076 &            \\
   17 & 2320.72 &     0.83 &     0.03 &    33.0 &     \ion{H}{1}$\lambda$972 &     1.3866 &            \\
   18 & 2321.88 &     0.11 &     0.01 &     9.3 &     \ion{H}{1}$\lambda$931 &     1.4949 &            \\
   19 & 2322.47 &     0.13 &     0.02 &     6.9 &                   {\Lya} ? &     0.9104 &            \\
   20 & 2324.65 &     0.46 &     0.02 &    21.8 &     \ion{H}{1}$\lambda$950 &     1.4478 &            \\
   21 & 2325.78 &     0.06 &     0.01 &     5.5 &                   {\Lya} ? &     0.9132 &            \\
   22 & 2331.23 &     0.04 &     0.01 &     3.0 &     \ion{H}{1}$\lambda$972 &     1.3983 &            \\
   23 & 2332.47 &     0.37 &     0.02 &    16.4 &     \ion{H}{1}$\lambda$972 &     1.3983 &            \\
   24 & 2334.98 &     0.19 &     0.02 &    10.1 &                   {\Lya} ? &     0.9207 &            \\
   25 & 2339.13 &     0.40 &     0.03 &    12.7 &     \ion{H}{1}$\lambda$938 &     1.4949 &            \\
   26 & 2342.22 &     0.08 &     0.01 &     5.5 &                   {\Lya} ? &     0.9267 &            \\
   27 & 2343.96 &     0.77 &     0.02 &    35.2 &   \ion{Fe}{2}$\lambda$2344 &          0 &            \\
   28 & 2349.69 &     0.53 &     0.02 &    24.8 &   \ion{Fe}{2}$\lambda$2374 &          0 &            \\
   29 & 2352.69 &     0.14 &     0.02 &     9.0 &                   {\Lya} ? &     0.9353 &            \\
   30 & 2353.30 &     0.08 &     0.02 &     5.2 &     \ion{H}{1}$\lambda$972 &     1.4242 &            \\
   31 & 2357.58 &     0.38 &     0.02 &    16.9 &                   {\Lya} ? &     0.9393 &            \\
   32 & 2359.25 &     0.34 &     0.03 &    13.5 &                   {\Lya} ? &     0.9407 &            \\
   33 & 2369.23 &     0.35 &     0.02 &    18.1 &     \ion{H}{1}$\lambda$950 &     1.4949 & Blend to the red with \ion{C}{3}$\lambda$977 at $z$=1.4242 \\
   34 & 2374.49 &     0.99 &     0.03 &    31.7 &   \ion{Fe}{2}$\lambda$2374 &          0 &            \\
   35 & 2380.41 &     0.66 &     0.02 &    27.7 &     \ion{H}{1}$\lambda$972 &     1.4478 &            \\
 \hline
 \end{tabular}
\end{table*}


\begin{table*}
 \centering
 \caption{Detection Rate of Important Transitions}\label{tab:tab4}
 \begin{tabular}{cccc}
 \hline
 \hline
 Transition                 & \#$_{cov}$$^{a}$ & \#$_{det}$$^{b}$ & Detection Rate$^c$ \\
                            &                  &                  &      (\%)      \\
 \hline
 \ion{C}{2}$\lambda$1335    &        27        &        15        &       56       \\
 \ion{C}{3}$\lambda$977     &        26        &        13        &       50       \\
 \ion{C}{4}$\lambda$1548    &        26        &        23        &       88       \\
 \ion{N}{2}$\lambda$1084    &        31        &         1        &        3       \\
 \ion{N}{5}$\lambda$1239    &        37        &         9        &       24       \\
 \ion{O}{6}$\lambda$1032    &        30        &        27        &       90       \\
 \ion{Al}{2}$\lambda$1671   &        18        &         7        &       39       \\
 \ion{Al}{3}$\lambda$1855   &        10        &         2        &       20       \\
 \ion{Si}{2}$\lambda$1260   &        36        &        13        &       36       \\
 \ion{Si}{3}$\lambda$1207   &        36        &        21        &       58       \\
 \ion{Si}{4}$\lambda$1394   &        26        &        13        &       50       \\
 \ion{Fe}{2}$\lambda$1608   &        22        &         2        &        9       \\
 \hline
 \end{tabular}
 \\
 $^a$ Number of systems in which transition is covered,
 $^b$ Number of systems in which transition is detected,
 $^c$ Probability that transition is detected at $>5\sigma$ for system in which it is covered.
\end{table*}

\clearpage


\renewcommand{\thefigure}{\arabic{figure}\alph{subfigure}}

\setcounter{subfigure}{1}
\begin{figure*}
\centering
\includegraphics[height=20cm]{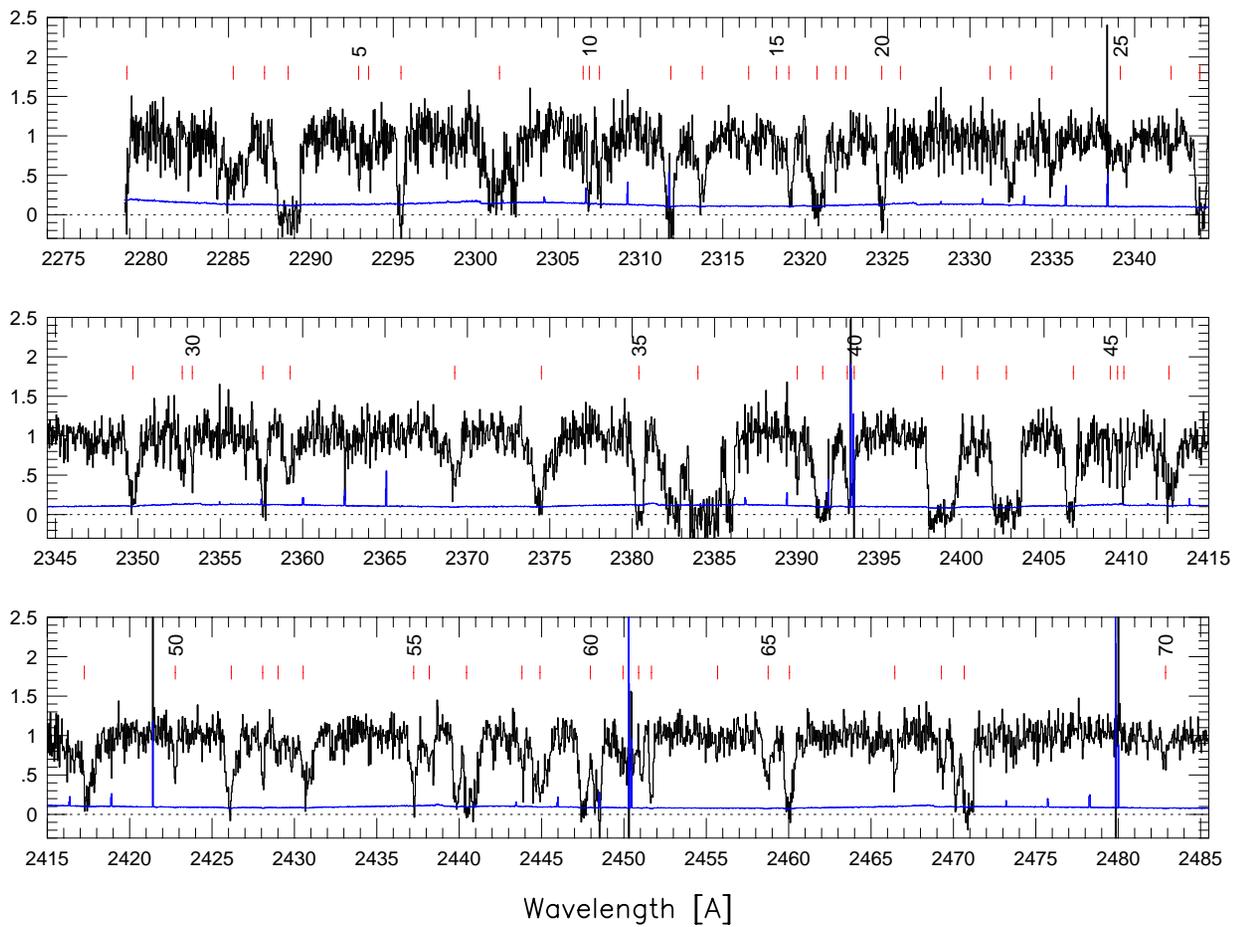}
\caption{Sample -- Normalized flux versus wavelength for a portion of
 the {\it HST}/E230M spectrum of quasar PG$0117+213$.  This is an
 example of a set of figures for all quasars and all wavelength
 coverage, which are given in the electronic version.  The histogram
 displayed beneath the data represents the error spectrum.  For
 reference, a dotted line is drawn at zero flux.  Numbered ticks mark
 spectral features detected at a $5\sigma$ level.  These are listed in
 Table~\ref{tab:tab3}.}
\end{figure*}
\clearpage

\addtocounter{figure}{7}
\setcounter{subfigure}{0}
\begin{figure*}
\epsfig{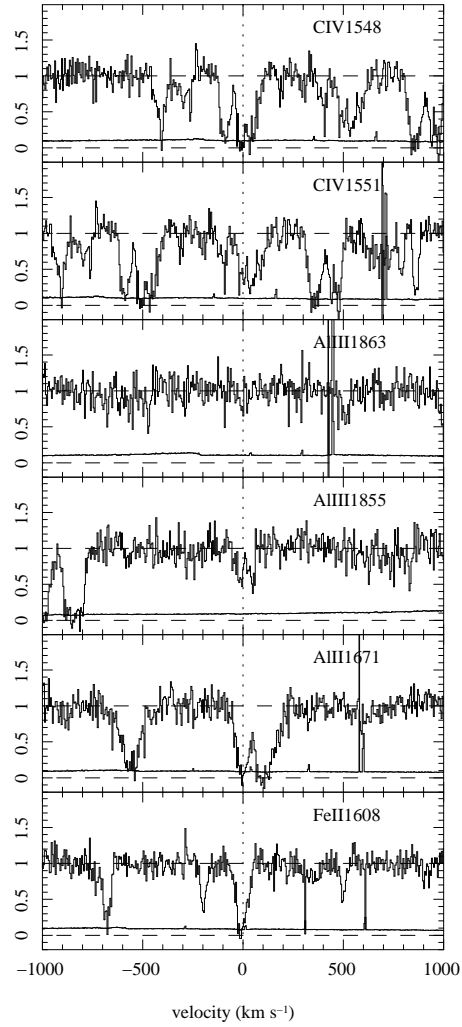}
\caption{Sample -- System plot for the $z=0.5764$ system toward
 PG0117$+$213. Important transitions are plotted, unless they are so
 badly compromised by a blend that no useful constraints can be
 gathered.  The transitions are aligned in velocity space, with
 $0$~{\kms} corresponding to $z$=0.5764. The error spectrum is plotted
 beneath the data, and a dashed line shown at zero flux.}
\end{figure*}


\begin{thebibliography}{99}
 \bibitem[\protect\citeauthoryear{Bahcall {\etal}}{1996}]{bah96}
  Bahcall J.~N., et al., 1996, ApJ, 457, 19
 \bibitem[\protect\citeauthoryear{Bahcall {\etal}}{1993}]{bah93}
  Bahcall J.~N., et al., 1993, ApJS, 87, 1
 \bibitem[\protect\citeauthoryear{Brown {\etal}}{2002}]{bro02}
  Brown, T., {\etal} 2002, {\it HST} STIS Data Handbook, version 4.0,
  ed. B. Mobasher (Baltimore; STScI)
 \bibitem[\protect\citeauthoryear{Charlton {\etal}}{2003}]{cha03}
  Charlton, J.~C., Ding,~J., Zonak, S.~G., Churchill, C.~W., Bond,
  N.~A., \& Rigby, J.~R. 2003, ApJ, 589, 311
 \bibitem[\protect\citeauthoryear{Churchill {\etal}}{2000}]{archive1}
  Churchill, C.~W., Mellon, R.~R., Charlton, J.~C., Jannuzi, B.~T.,
  Kirhakos, S., Steidel, C.~C., \& Schneider, D.~P. 2000, ApJS, 130,
  91
 \bibitem[\protect\citeauthoryear{Churchill {\etal}}{1999}]{chu99}
  Churchill, C.~W., Rigby, J.~R., Charlton, J.~C., \& Vogt,
  S.~S. 1999, ApJS, 120, 51
 \bibitem[\protect\citeauthoryear{Churchill \& Vogt}{2001}]{chu01}
  Churchill, C.~W., \& Vogt, S.~S. 2001, AJ, 122, 679
 \bibitem[\protect\citeauthoryear{Ding {\etal}}{2005}]{din05} 
  Ding, J., Charlton, J.~C., \& Churchill, C.~W. 2005, ApJ, 621, 615
 \bibitem[\protect\citeauthoryear{Ding {\etal}}{2003}]{din03a} 
  Ding, J., Charlton, J.~C., Churchill, C.~W., \& Palma, C. 2003a, ApJ,
  590, 746
 \bibitem[\protect\citeauthoryear{Ding {\etal}}{2003}]{din03b} 
  Ding J., Charlton J.~C., Bond N.~A., Zonak S.~G., Churchill C.~W.,
  2003b, ApJ, 587, 551
 \bibitem[\protect\citeauthoryear{Ganguly {\etal}}{2007, in
  preparation}]{gan07} Ganguly R. et al., 2007, in preparation
 \bibitem[\protect\citeauthoryear{Jannuzi {\etal}}{1998}]{jan98}
  Jannuzi B.~T., et al., 1998, ApJS, 118, 1
 \bibitem[\protect\citeauthoryear{Kirkman {\etal}}{2007}]{kir07}
  Kirkman, D., Tytler, D., Lubin, D., \& Charlton, J.\ 2007,
  MNRAS, 376, 1227
 \bibitem[\protect\citeauthoryear{Kirkman {\etal}}{2001}]{kir01}
  Kirkman, D., et al.\ 2001, ApJ, 559, 23
 \bibitem[\protect\citeauthoryear{Koratkar {\etal}}{1998}]{kor98}
  Koratkar A., Antonucci R., Goodrich R., Storrs A., 1998, ApJ, 503,
  599
 \bibitem[\protect\citeauthoryear{Levshakov {\etal}}{2003}]{lev03} 
  Levshakov S.~A., Agafonova I.~I., Reimers D., Baade R., 2003, A\&A, 
  404, 449
 \bibitem[\protect\citeauthoryear{Masiero {\etal}}{2005}]{mas05}
  Masiero, J.~R., Charlton, J.~C., Ding, J., \& Churchill, C.~W. 2005,
  ApJ, 623, 57
 \bibitem[\protect\citeauthoryear{Milutinovi{\'c} {\etal}}{2006}]{mil06} 
  Milutinovi{\'c} N., Rigby J.~R., Masiero J.~R., Lynch R.~S., Palma C., 
  Charlton J.~C., 2006, ApJ, 641, 190
 \bibitem[\protect\citeauthoryear{Narayanan {\etal}}{2005}]{nar05}
  Narayanan, A., Charlton, J.~C., Masiero, J.~R., \& Lynch, R., 2005,
  ApJ, 632, 92
 \bibitem[\protect\citeauthoryear{Rao \& Turnshek}{2000}]{rao00}
  Rao, S.~M., \& Turnshek, D.~A., 2000, ApJS, 130, 1
 \bibitem[\protect\citeauthoryear{Reimers et al.}{2003}]{rei03}
  Reimers D., Baade R., Quast R., Levshakov S.~A., 2003, A\&A, 410,
  785
 \bibitem[\protect\citeauthoryear{Savage {\etal}}{2002}]{savage}
  Savage, B.~D., Sembach, K.~R., Tripp, T.~M., \& Richter, P., 2002, ApJ,
  564, 631
 \bibitem[\protect\citeauthoryear{Schneider {\etal}}{1993}]{sch93} 
  Schneider, D.~P., {\etal}\ 1993, ApJS, 87, 45
 \bibitem[\protect\citeauthoryear{Tripp {\etal}}{2000}]{tripp}
  Tripp, T.~M., Savage, B.~D., \& Jenkins, E.~B. 2000, ApJ, 534, L1
 \bibitem[\protect\citeauthoryear{Verner {\etal}}{1996}]{vvf96}
  Verner, D.~A., Verner, E.~M., \& Ferland, G.~J. 1996, ADNDT, 64, 1
 \bibitem[\protect\citeauthoryear{Zonak {\etal}}{2004}]{zon04} 
  Zonak, S.  G., Charlton, J. C., Ding, J. \& Churchill, C. W. 2004,
  ApJ, 606, 196
\end{thebibliography}
\end{document}